\documentclass[twocolumn,notitlepage,prl,superscriptaddress]{revtex4-1}
\usepackage[usenames,dvipsnames]{color}
\usepackage{amsmath}
\usepackage{amssymb}
\usepackage{graphicx}
\usepackage{tabularx}
\newcommand*\Eval[3]{\left.#1\right\rvert_{#2}^{#3}}
\begin{document}
\title{Superconducting Fluctuations in the Normal State of the Two-Dimensional Hubbard Model}

\author{Xi Chen}
\affiliation{Department of Physics, University of Michigan, Ann Arbor, Michigan 48109, USA} 
\author{J. P. F. LeBlanc}
\affiliation{Department of Physics, University of Michigan, Ann Arbor, Michigan 48109, USA} 
\author{Emanuel Gull}
\affiliation{Department of Physics, University of Michigan, Ann Arbor, Michigan 48109, USA}

\date{\today}

\begin{abstract}
We compute the two-particle quantities relevant for superconducting correlations in the two-dimensional Hubbard model within the dynamical cluster approximation. In the normal state we identify the parameter regime in density, interaction, and second-nearest-neighbor hopping strength that maximizes the $d_{x^2-y^2}$ superconducting transition temperature.  We find in all cases that the optimal transition temperature occurs at intermediate coupling strength, and is suppressed at strong and weak interaction strengths. Similarly, superconducting fluctuations are strongest at intermediate doping and suppressed towards large doping and half-filling.
We find a change in sign of the vertex contributions to $d_{xy}$ superconductivity from repulsive near half filling to attractive at large doping. $p$-wave superconductivity is not found at the parameters we study, and $s$-wave contributions are always repulsive. For negative second-nearest-neighbor hopping the optimal transition temperature shifts towards the electron-doped side in opposition to the van Hove singularity which moves towards hole doping.  We surmise that an increase of the local interaction of the electron-doped compounds would increase $T_c$.
\end{abstract}


\maketitle

Understanding physical scenarios that give rise to superconductivity at high temperatures has been a primary motivating force behind computational research of strongly correlated electron systems and candidate models such as the 2D Hubbard model \cite{Anderson87,scalapino:2007}.  Only recently have reliable many-body methods \cite{benchmark:2015} become powerful enough to reach temperatures low enough to cross the superconducting transition at intermediate interaction strengths \cite{Scalapino12,gull:2013,gull:2012}, but progress is limited by the exponential scaling intrinsic to all unbiased methods. Such computational work has identified clearly the competition between correlations that give rise to superconductivity and other phases such as antiferromagnetism \cite{Zheng15,Carlos15} and the pseudogap \cite{gull:2013,gull:2012} phenomenon within the 2D Hubbard model.

Central to understanding these phases is the evaluation of two-particle susceptibilities and vertex functions at nonzero temperature, which diverge on approach to a continuous phase transition and may also exhibit signs of a transition at temperatures much larger than the transition temperature, for parameters that are accessible with current techniques and computational power. 
Nevertheless, the numerical calculation of these two-particle susceptibilities requires techniques that are robust across the full phase diagram, can reach low temperatures, are capable of providing reliable and systematically improvable results, and are able to distinguish independent phases. 
Cluster dynamical mean field theory \cite{hettler:1998,hettler:2000,Lichtenstein00,Kotliar01,Maier05} provides such a self consistent non-perturbative tool for simulating strongly correlated electron problems. The dynamical cluster approximation (DCA) is based on a self-energy discretization into $N_c$ independent self-energy coefficients which recover the exact limit as $N_c \to \infty$ \cite{fuchs:2011,leblanc:2013,Staar13,benchmark:2015} and capture much of the physics believed to be relevant for the superconductivity and pseudogap physics two-dimensional Hubbard model on clusters of size $8$ and larger \cite{gull:2013,gull:2012}.

In this work, we specifically address the problem of optimizing the superconducting transition temperature in the 2D Hubbard model by analyzing wide regions of parameter space.  
We first demonstrate how the vertex contribution to the pairing susceptibility can be used as an indicator of the proximity to the superconducting transition temperature, $T_c$. We then show that this quantity, as temperature is reduced, mimics the dependence of $T_c$ on model parameters.  
This allows us to sweep the entirety of parameter space in density $n$, interaction strength $U/t$, and second-nearest neighbor hopping $t'/t$ at numerically accessible $T \sim 2T_c$, to identify regions of qualitatively high or low $T_c$, so that the maxima can then be targeted for a quantitative determination of the optimal $T_c$ value.  
We mainly focus on $d_{x^2-y^2}$ superconductivity but show that $d_{xy}$ superconducting fluctuations change from attractive (at large doping) to repulsive (at low doping), $p$-wave fluctuations are always either repulsive or zero within error bars at the system sizes, interaction strengths, and dopings we study, and $s$ wave contributions are strongly repulsive.

We study the single orbital Hubbard model in two dimensions with nearest and next-nearest hopping parameters,
\begin{equation}
H = \sum_{k,\sigma} \left(\epsilon_{k} -\mu\right)c_{k\sigma}^\dagger c_{k\sigma}+U\sum_i n_{i\uparrow}n_{i\downarrow},
\label{H}
\end{equation}
where $\mu$ is the chemical potential, $k$ momentum, $i$ labels sites in real-space, $U$ is the interaction, and the dispersion is given by
$\epsilon_k=-2t\left[\cos(k_x)+\cos(k_y)\right]-4t'\cos(k_x)\cos(k_y).$
We operate in a formalism that allows for a nonzero anomalous Green's function in the superconducting state, which is defined as
$F(k,\tau)=-\langle T_\tau c_{k\uparrow}(\tau)c_{-k\downarrow}(0)\rangle$.
At $T>T_c$ superconducting order will be absent but fluctuations are captured by the generalized susceptibility, written in imaginary time in terms of the one- and two-particle Green's functions as \cite{Rohringer12} (see Supplemental Material \cite{Supplemental} for definition and notations)
\begin{align}
\chi_{\sigma_1\sigma_2\sigma_3\sigma_4}(k_1\tau_1,k_2\tau_2,k_3\tau_3,k_4\tau_4)\\=G_{2,\sigma_1...\sigma_4}(k_1\tau_1,k_2\tau_2,k_3\tau_3,k_4\tau_4)\nonumber\\ \nonumber-G_{\sigma_1\sigma_2}(k_1\tau_1,k_2\tau_2)G_{\sigma_3\sigma_4}(k_3\tau_3,k_4\tau_4)
\end{align}
or as its Fourier transform
\begin{align}
\chi_{pp\sigma\sigma'}^{\omega\omega'\nu}(k,k',q) = \int_{0}^\beta\int_{0}^\beta\int_{0}^\beta d\tau_1d\tau_2d\tau_3
\\ \nonumber\times \chi_{\sigma\sigma\sigma'\sigma'}(k\tau_1,(q-k')\tau_2,(q-k)\tau_3,k'0)\\ \nonumber\times e^{-i\omega\tau_1}e^{i(\nu-\omega')\tau_2}e^{-i(\nu-\omega)\tau_3}
\end{align}
where $\omega$ and $\omega'$ are fermionic Matsubara frequencies, $\nu$ is a bosonic Matsubara frequency, $\sigma$ and $\sigma'$ are $\uparrow$ or $\downarrow$ spin labels and $k$, $k'$ and $q$ are initial, final and transfer momenta respectively, and $pp$ denotes the Fourier transform convention.  With the difference between the $\sigma \sigma' \equiv\uparrow \uparrow$ and $\uparrow \downarrow$ susceptibilities defined as
$\chi_{pp\overline{\uparrow\downarrow}}=\chi_{pp\uparrow\uparrow}-\chi_{pp\uparrow\downarrow},$
linear response theory relates $\chi_{pp\overline{\uparrow\downarrow}}$ to the response of a system to a generating superconducting field $\eta(k)$
\begin{equation}\label{eqn:linear_response_first}
\int_0^\beta d\tau \Eval{\frac{\delta F(k',\tau=0;\eta)}{\delta\eta(k,\tau)}}{\eta=0}{ }\!\!\!\!\!=\frac{1}{\beta^2}\sum\limits_{\omega\omega'}\chi_{pp\overline{\uparrow\downarrow}}^{\omega\omega'\nu=0}(k,k',q=0)
\end{equation}
where $F(k',\tau;\eta)$ is the anomalous Green's function computed in the presence of an external superconducting field.
We note that the quantity on the left-hand side is commonly referred to as the uniform pairing susceptibility \cite{White89B,Khatami15}.

Continuous phase transitions can be identified by the point in phase space where the corresponding susceptibility diverges. The susceptibility can then, using the Bethe-Salpeter equation, be separated into a `bare' contribution
\begin{align}
\chi_{0pp}^{\omega\omega'\nu}(k,k',q) = -\beta G_\sigma(k,i\omega)G_\sigma(q-k',i\nu-i\omega')\delta_{\omega\omega'}\delta_{kk'}.
\end{align}
which never diverges and a part containing an irreducible vertex function $\Gamma_{pp}$,
\begin{align}\label{Bethe_k}
\chi_{pp\overline{\uparrow\downarrow}}^{\omega\omega'\nu}(k,k',q)=\chi_{0pp}^{\omega\omega'\nu}(k,k',q)-\frac{1}{\beta^2}\chi_{pp\overline{\uparrow\downarrow}}^{\omega\omega''\nu}(k,k'',q)  \nonumber\\ \times \Gamma_{pp\overline{\uparrow\downarrow}}^{\omega''\omega'''\nu}(k'',k''',q)\chi_{0pp}^{\omega'''\omega'\nu}(k''',k',q).
\end{align}
In order to see the origin of the divergence in $\chi_{pp\overline{\uparrow\downarrow}}^{\omega\omega'\nu}$ this susceptibility can be expressed in matrix notation giving
\begin{align}
{\chi}_{pp\overline{\uparrow\downarrow}}=\frac{{\chi}_0}{1+\frac{1}{\beta^2}\Gamma_{pp\overline{\uparrow\downarrow}}{\chi}_0}.
\end{align}
and the point of divergence of $\chi$ is identified as the point where an eigenvalue of $-\frac{1}{\beta^2}\Gamma_{pp\overline{\uparrow\downarrow}}{\chi}_0$ crosses $1$, and the symmetry of the eigenvector will identify the symmetry of the order parameter.

\begin{figure}
\includegraphics[width=80mm]{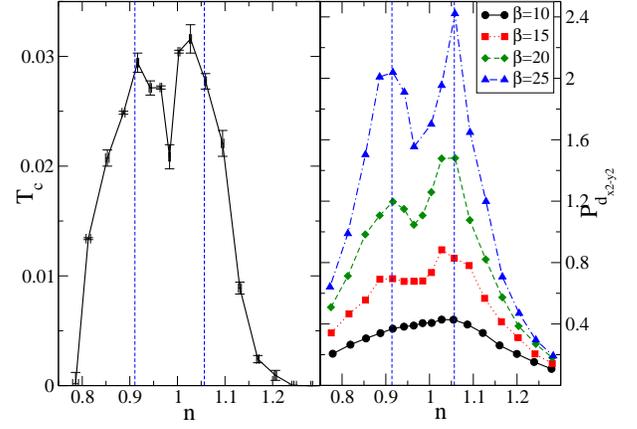}
\caption{Left panel: Superconducting critical temperature of the Hubbard model with nearest neighbour hopping and next nearest neighbour hopping $t'=-0.1t$ for $U=6t$ using an $N_c=8$ dynamical cluster approximation. Right panel: $P_{d_{x^2-y^2}}$ at different temperatures with interaction strength $U=6t$ and next nearest neighbour hopping $t'=-0.1t$ using an $N_c=8$ cluster.\label{fig:TcAndChiMinusChi0Comparison}}
\end{figure}
\begin{figure*}[bth]
\includegraphics[width=\textwidth]{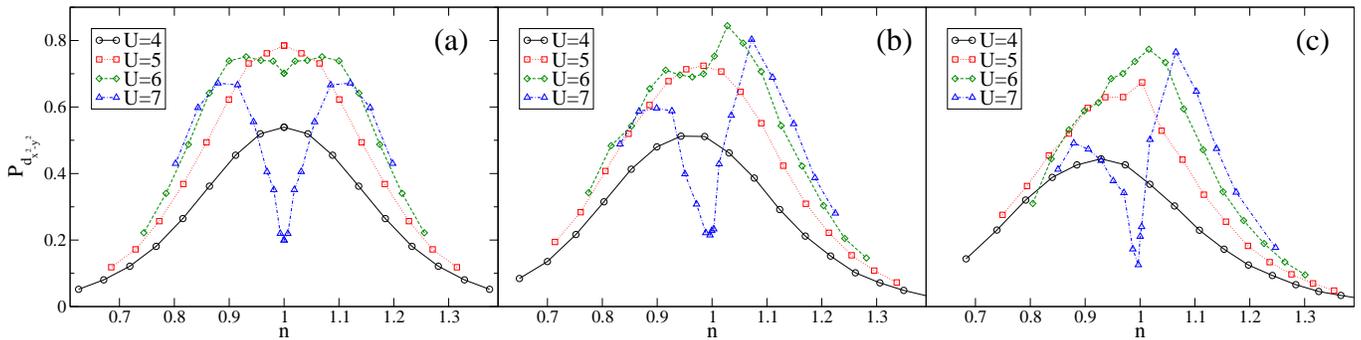}
\caption{$P_{d_{x^2-y^2}}$ for different interaction strengths as a function of carrier concentration on an eight-site cluster, for $t'=0$ (panel a), $t'=-0.1t$ (panel b) and $t'=-0.2t$ (panel c) at $\beta=15/t$. $U=4t$ (black solid line, circles), $5t$ (red dotted line, squares), $6t$ (green dashed line, diamonds) and $7t$ (blue dash-dotted line, triangles).\label{fig:ChiMinusChi0Tprimes}}
\end{figure*}

In what follows we solve the Hubbard model within the (paramagnetic) dynamical cluster approximation which  approximates the self-energy of the interacting model by a number, $N_c$, of `coarse-grained' frequency-dependent but momentum-independent self-energy tiles.
We primarily present results for an $N_c=8$ cluster since this is the smallest DCA system that captures a clear distinction between nodal and antinodal physics \cite{Werner098site,Gull09_8site,Lin10,Gull10_clustercompare}. Comparisons to larger and smaller $N_c=4$ and $N_c=16$ systems are shown in the supplemental materials \cite{Supplemental}. Antiferromagnetic order is actively suppressed in our calculations by enforcing paramagnetic spin symmetry, and the presence or effect of charge order \cite{Misawa14} has not been investigated.
The DCA calculation provides one- and two-particle cluster Green's functions, from which we extract cluster susceptibilities and, using the formalism outlined in Ref.~\cite{Fotso}, the coarse-grained lattice susceptibilities $\chi_{pp\overline{\uparrow\downarrow}}^{\omega\omega'\nu}(K,K',Q)$, where $K$, $K'$ and $Q$ are cluster momenta. 
In order to analyze the angular dependence of the superconducting order, one typically performs a multipole expansion restricted to the $D_{4h}$ square lattice symmetry.\cite{allen:1976,sigrist:1991,chen:1993} Because of our limited momentum resolution we project out and analyze the leading contribution and are insensitive to higher order harmonics around the Fermi surface.
The accessible $s-,\ p-, d_{xy}$ or $d_{x^2-y^2}$ symmetries are enforced by including symmetry factors $g(K)g(K')$ while summing over all initial $K$ and final $K'$ states in Eq.~(\ref{eqn:linear_response_first}) \cite{White89,White89B,Deng14}, with
\begin{equation}\label{eqn:symmetry_defs}
g(K) = \left\{
  \begin{array}{ll}
    1 &  s\\
    \sin(K_x) &  p\\
    \sin(K_x)\sin(K_y) &  d_{xy}\\
    \cos(K_x)-\cos(K_y) &  d_{x^2-y^2}
  \end{array}
\right. .
\end{equation}
The divergence of $\chi^{\omega\omega'\nu}$ is caused by the vertex correction part $\chi^{\omega\omega'\nu} - \chi_0^{\omega\omega'\nu}$.  We impose a shorthand notation for this quantity of interest, which we call the correlated pairing susceptibility $P_g$, where $g$ refers to the corresponding symmetry function defined in Eq.~\ref{eqn:symmetry_defs}, and we take this to be the summation over fermionic Matsubara frequencies and momenta:
\begin{align}
P_g := &(\chi - \chi_0)_g = \frac{1}{\beta^2}\sum\limits_{\omega\omega' K K'}g(K)g(K')\\ \nonumber \times &\left[\chi_{pp\overline{\uparrow\downarrow}}^{\omega\omega'0}(K,K',0) - \chi_{0}^{\omega\omega'0}(K,K',0) \right]/\sum\limits_K g(K)^2.
\end{align}
We show in the supplemental material \cite{Supplemental} an explicit example where the point of divergence of $\chi$ coincides with the divergence of a single  eigenvalue with $d_{x^2-y^2}$ symmetry.

The fact that the correlated pairing susceptibility $P_g$ must become large on approach to $T_c$ grants us additional insights at $T>T_c$, where $P_g$ can be used as a qualitative measure of the proximity of the system to a transition. 
The left panel of Fig.~\ref{fig:TcAndChiMinusChi0Comparison} shows the critical temperature obtained from systematically reducing $T$ and explicitly evaluating the eigenvalues of $-\frac{1}{\beta^2}\Gamma_{pp\overline{\uparrow\downarrow}}{\chi}_0$ to find the divergence of the $d_{x^2-y^2}$ susceptibility.
The right panel contrasts this with the magnitude of $P_g$ at much higher temperatures $\beta$ $=$ $10/t$, $15/t$, $20/t$, and $25/t$.
We see $P_g$ tracks $T_c$ and shows the largest superconducting fluctuations approximately where $T_c$ is highest, as also indicated by the vertical blue lines.
The correspondence of $P_g$ to $T_c$ improves as $T$ decreases towards $T_c$.

In Fig.~\ref{fig:ChiMinusChi0Tprimes}(a) we explore $P_{d_{x^2-y^2}}$ as a function of particle density $n$ ($n=1$ denotes half filling) in the intermediate interaction strengths regime $U/t=4$ to $7$ at $\beta = 15/t\approx 2T_c$. 
For the weakest interaction strength considered here, $U=4t$, the  superconducting $d_{x^2-y^2}$ fluctuations are strongest at half filling and decrease rapidly towards larger hole and electron doping.
At $10\%$ doping, the model has been shown to be $d_{x^2-y^2}$ superconducting by DCA calculations extrapolated to the thermodynamic limit \cite{Maier05B}, and $8$-site fluctuations have shown to be weaker than for the lattice model.
The maximum of fluctuations at half filling is consistent with results from weak coupling theory \cite{Scalapino10}, FLEX \cite{Arita00}, and diagrammatic Monte Carlo calculations in the weak coupling limit \cite{Deng14}, and is also observed in results from lattice quantum Monte Carlo (QMC) simulation \cite{Santos89} and the two-particle-self-consistent approximation \cite{Tremblay12}. Reduction of $U$ rapidly suppresses the strength of fluctuations.
$P_{d_{x^2-y^2}}$ increases at all $n$ as $U$ is raised to $5t$.
As $U$ is further raised to $6t$, the strength of fluctuations increases away from half filling but decreases near half filling, and the fluctuation maximum moves to finite doping, establishing a dome. The suppression at half filling coincides with the establishment of a pseudogap at this interaction strength  \cite{Werner098site,Gull09_8site}, and is also seen in QMC simulation \cite{White89B} and TPSC \cite{Kyung03,Tremblay12} (though it seems to be absent in four-site CDMFT \cite{Sordi12}). Simulations directly in the superconducting phase \cite{gull:2012} have also shown that that superconductivity in this region is suppressed.
Above $U/t=6.4$ the half-filled system becomes Mott insulating \cite{Gull09_8site} and superconducting fluctuations are further suppressed (but remain nonzero), while their maximum strength moves to higher doping, giving the appearance of a dome structure centered at doping, $\delta\sim\pm 1/8$ for $U/t\sim 8$.
As the interaction strength is further increased, fluctuations are suppressed and quickly decay, in qualitative agreement with simulations of the $t-J$ model \cite{Martins01} and Hubbard NLCE calculations \cite{Khatami15}.

\begin{figure}
\includegraphics[width=\columnwidth]{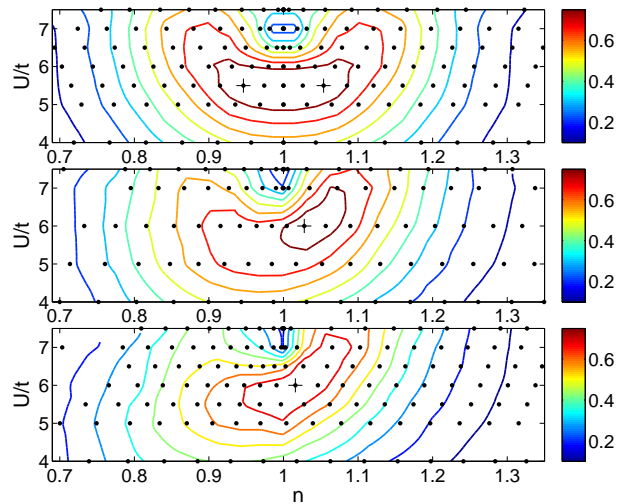}
\caption{Contour plots for $P_{d_{x^2-y^2}}$ in space of interaction strength and carrier concentration on an 8-site cluster at $\beta=15/t$. Top panel:  $t'=0$. Middle panel: $t'=-0.1t$. Bottom panel: $t'=-0.2t$. $T_c^{max}$ occurs at $(U^{max}, n^{max})=(5.5,0.95 \text{ and } 1.05)$, $(6,1.03)$, $(6,1.01)$ respectively, marked by a + symbol.\label{fig:contour}}
\end{figure}

Figure \ref{fig:contour} expands further upon the data of Fig.~\ref{fig:ChiMinusChi0Tprimes}, including additional data points at intermediate interaction values, as a false color contour plot at $t'=0$ in Fig.~\ref{fig:contour}(a). The plot clearly shows the intermediate interaction region most conducive to superconductivity.  The point of maximum susceptibility which occurs at ${U^{max},n^{max}}$,  is marked as + and occurs at $U/t=6$, $n=0.92$ for the eight-site cluster.  A wide area in the vicinity of this point exhibits fluctuation within $10\%$ of the maximum value, showing that $d_{x^2-y^2}$ superconducting fluctuation is a robust feature of the model. Finite size effects change the precise location and general strength of the fluctuations (see Supplemental Material \cite{Supplemental}) but not the overall shape. Long-range antiferromagnetism may preempt the superconducting phase near half filling; see e.g. Ref.~\cite{Zheng15}.

Next-nearest neighbor hopping, shown in Figs.~\ref{fig:ChiMinusChi0Tprimes}(b), \ref{fig:ChiMinusChi0Tprimes}(c) and Fig.~\ref{fig:contour}, has a profound effect on $d_{x^2-y^2}$ fluctuations.
As the interaction strength is raised, a pronounced particle hole asymmetry appears for $t'/t=-0.1$ (panel (b)) that increases superconducting fluctuations on the electron doped side ($n>1$) while suppressing them on the hole doped side. 
Increasing $t'$ to $-0.2t$ (Fig.~\ref{fig:ChiMinusChi0Tprimes}(c)) leads to a further enhancement of fluctuations on the electron doped side and increased suppression on the hole doped side near half filling.
 This behavior seems to be unrelated to any feature in the single particle density of states which has a van Hove maximum on the hole-doped side. Rather, we attribute it to the establishment of a pseudogap on the hole doped side, which is absent on the electron doped side \cite{Gull09_8site}, and which is known to rapidly suppress critical temperature near half filling \cite{gull:2012}. The magnitude of fluctuation at the electron doped side (and outside of the pseudogap region at the hole-doped side) is not significantly changed, suggesting (in agreement with ED and DMRG simulations on $t-J$ ladders \cite{White99,Martins01} and NCA results on $2\times 2$ clusters) that the $t'$ trends observed in real materials are not captured by the single band Hubbard model \cite{Pavarini01}. We find that further increase of $t'$ continues this trend and reduces the overall susceptibility to $d_{x^2-y^2}$ superconductivity.
 
 Our results suggest that the low-energy effective models of high $T_c$ compounds do not just differ by $t'$, but also by their on-site interactions $U$. As the electron-doped compounds have a much lower critical temperature than the hole doped ones, we surmise that they are not localized at the point in phase space that yields the highest $T_c$, and that an increase of $U$ would rapidly increase the critical temperature.

\begin{figure}
\includegraphics[width=\columnwidth]{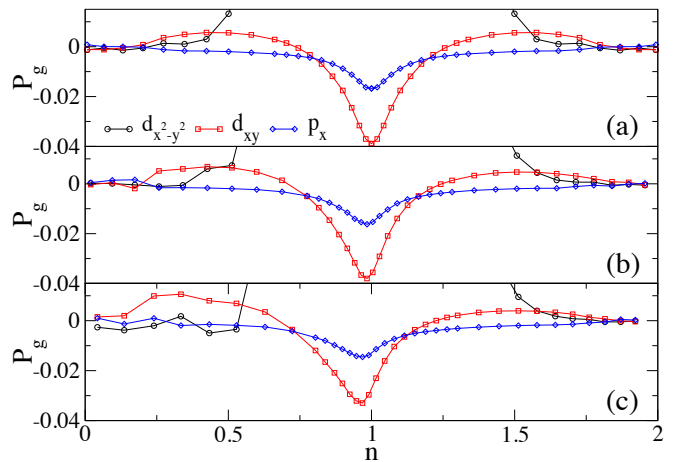}
\caption{\label{fig:pd_sym}The correlated pairing susceptibility $P_g$ in different symmetry channels with interaction strength $U=6t$, at $\beta t=15$, using 8-site cluster. Panel(a): $t'=0$; panel(b): $t'=-0.1t$; panel(c): $t'=-0.2t$}
\end{figure}

%
Finally, we establish the absence of high-temperature superconducting fluctuations in other symmetry channels by considering $g(k)g(k')$ factors with alternate symmetry in the two-particle representation of the susceptibility.
We plot results for $t'/t=0$, -0.1, and -0.2 in Figs.~\ref{fig:pd_sym}(a$\to$c) at $U/t=6$, for $d_{xy}$ and $p$-wave symmetry and include $d_{x^2-y^2}$ for reference (also shown in Fig.~\ref{fig:ChiMinusChi0Tprimes}).  

In the large doping weak coupling regime, $d_{x^2-y^2}$ superconductivity is preempted by $d_{xy}$ superconductivity \cite{chubukov:1992, baranov:1992}. This is also found in RPA calculations \cite{fukazawa:2001,Romer15} and diagrammatic QMC calculations \cite{Deng14}. In contrast, the vertex contribution to $d_{xy}$ superconductivity is repulsive near half filling, consistent with early QMC calculations \cite{White89B}. Figure~\ref{fig:pd_sym} shows  how it changes sign for larger doping and eventually becomes the dominant contribution.

As $U$ is raised in the dilute ($n\rightarrow 0$) limit, $d_{xy}$ order is replaced immediately by $p$-wave superconductivity \cite{Arita00,Deng14}.
Third order perturbative calculations \cite{fukazawa:2001} also find a large range of $p$-wave stability (but no $d_{xy}$) in the large doping regime at $U=6$, and DCA calculations similarly found dominant $p$-wave contributions \cite{arita:2006}. Within our calculations, $p$-wave contributions to the vertex are zero within errors in the entire range of phase space, except near half filling, where they are repulsive. Our data are consistent with Ref.~\cite{arita:2006} on the level of the susceptibility, but we find that the dominant contribution observed in that work is carried by $\chi_0$, not the vertex part. Whether a DCA simulation could find dominant $p$-wave contributions to the vertex at smaller $U$, lower $T$, or on larger systems is an open question. The highest critical temperature of any non-$d_{x^2-y^2}$ superconductivity is far below the $T$ examined in this work.

Over the entire phase space, $s$-wave superconductivity (not plotted in Fig.~\ref{fig:pd_sym}) is strongly repulsive, consistent with QMC calculations \cite{White89B,Scalapino93,Tremblay94}.
At $t'=-0.2$ and in the dilute limit, weak coupling and RPA results suggest a favored $d_{xy}$ symmetry \cite{chubukov:1992,chubukov:1993,Romer15}, consistent with our results at larger $U/t$ and high temperatures.

In summary, we have identified the regions in parameter space that give optimal superconducting transition temperatures, using a formalism based on two-particle simulations at temperatures much higher than $T_c$, We have explored the susceptibility of the Hubbard model towards superconducting order over the entirety of the phase diagram.

We find that both weak and strong interaction regimes, as well as low doping and half filled regimes, are nonoptimal for superconducting fluctuations, but that there is a large region that is very conducive to superconductivity. For $t'<0$ we find a shift of the optimal superconducting features to the electron-doped side of the phase diagram, due to the establishment of a competing pseudogap on the hole-doped side. As actual electron-doped compounds have a lower $T_c$ than the hole-doped ones, we surmise that a rapid increase of $T_c$ could be achieved by changing the effective on-site interaction.

By examining alternate order symmetries and $t'<0$ we show susceptibility towards $d_{xy}$ but not $p$-wave superconductivity in the strongly hole doped $n\to0$ (dilute) limit. We emphasize that transitions to those symmetries happen at temperatures much lower than the $T$ we have examined here.

\begin{acknowledgments}
This project was supported by the Simons Foundation collaboration on the many-electron problem. We would like to thank Andrew J. Millis for insightful and helpful discussions. This research used resources of the National Energy Research Scientific Computing Center, a DOE Office of Science User Facility supported by the Office of Science of the U.S. Department of Energy under Contract No. DE-AC02-05CH11231. Our codes are based on the open source ALPS \cite{ALPS20} library.
\end{acknowledgments}

\bibliographystyle{apsrev4-1}
\bibliography{refs}
\end{document}